\documentclass[sigconf]{acmart}

\usepackage{soul}
\usepackage{amsmath}
\usepackage{subfig}
\usepackage{float}
\usepackage{array}
\usepackage{multirow}
\AtBeginDocument{%
  \providecommand\BibTeX{{%
    \normalfont B\kern-0.5em{\scshape i\kern-0.25em b}\kern-0.8em\TeX}}}

\copyrightyear{2022}
\acmYear{2022}
\setcopyright{acmcopyright}\acmConference[JCDL '22]{The ACM/IEEE Joint Conference on Digital Libraries in 2022}{June 20--24, 2022}{Cologne, Germany} \acmBooktitle{The ACM/IEEE Joint Conference on Digital Libraries in 2022 (JCDL '22), June 20--24, 2022, Cologne, Germany}
\acmPrice{15.00}
\acmDOI{10.1145/3529372.3530931}
\acmISBN{978-1-4503-9345-4/22/06}

\begin{document}

\title{Studying Retrievability of Publications and Datasets in an Integrated Retrieval System}

\author{Dwaipayan Roy}
\affiliation{%
  \institution{Indian Institute of Science Education and Research, Kolkata}
  \city{}
  \country{India}
}
\email{dwaipayan.roy@iiserkol.ac.in}

\author{Zeljko Carevic}
\affiliation{%
  \institution{GESIS -- Leibniz Institute for the Social Sciences, Cologne}
  \city{}
  \country{Germany}
}
\email{Zeljko.Carevic@gesis.org}

\author{Philipp Mayr}
\affiliation{%
  \institution{GESIS -- Leibniz Institute for the Social Sciences, Cologne}
  \city{}
  \country{Germany}
}
\email{Philipp.Mayr@gesis.org}

\renewcommand{\shortauthors}{D. Roy et al.}

\begin{abstract} In this paper, we investigate the retrievability of datasets and publications in a real-life Digital Library (DL). The measure of \emph{retrievability} was originally developed to quantify the influence that a retrieval system has on the access to information. Retrievability can also enable DL engineers to evaluate their search engine to determine the ease with which the content in the collection can be accessed. Following this methodology, in our study, we propose a system-oriented approach for studying dataset and publication retrieval. A speciality of this paper is the focus on measuring the accessibility biases of various types of DL items and including a metric of \emph{usefulness}. 
Among other metrics, we use Lorenz curves and Gini coefficients to visualize the differences of the two retrievable document types (specifically datasets and publications).
Empirical results reported in the paper show a distinguishable diversity in the retrievability scores among the documents of different types. 
\end{abstract}


\maketitle

\section{Introduction}

With the availability of various types of information from different sources, sometimes with different modalities, the necessity of federated or integrated search systems~\cite{adali:icde1995, federated-arguello} has become a prominent research topic in information retrieval and digital library.
Textual data still remains the predominant type among them and significant research has been conducted in the domain of textual document retrieval. 
Among the rest, recent research on dataset retrieval~\cite{dataset-ret} has become increasingly important in the (interactive) information retrieval and digital library community. One of the reasons is undoubtedly the massive amount of research datasets available. However, the underlying characteristics of dataset retrieval also contribute to the attention in this area. One often mentioned characteristic is the increased complexity of datasets over traditional document retrieval. While the latter is well known and adequately studied, datasets often include more extensive material and structures that are relevant for retrieval. This may involve the raw data, descriptions of how the data was collected, taxonomic information, questionnaires, codebooks, etc. 
Recently, numerous studies have been conducted to further identify the characteristics of dataset retrieval. These studies include the observation of data retrieval practices \cite{kramer2021data}, interviews and online questionnaires \cite{kern2015there,friedrich2020looking} and transaction log analysis \cite{kacprzak2017query,carevic2020characteristics}.

In this paper, we follow a system-oriented approach for studying dataset retrieval. By employing the measure of \em retrievability \em \cite{ret-cikm08}, we aim to gain insights into the particularities of dataset retrieval in comparison to traditional document retrieval. The measure of retrievability was originally developed to quantify the influence that a retrieval system has on the access to information. In a simplified way, retrievability represents the ease with which a document can be retrieved given a particular IR system \cite{ret-cikm08}. The measure of retrievability can be utilised for several use cases. Retrievability can, for instance, be employed to identify potential biases in the underlying corpus. Retrievability can also enable digital library (DL) engineers to evaluate their search engine to determine the ease with which the content in the collection can be accessed.  

In this paper, we investigate the retrievability of various types of documents in the integrated DL \textit{GESIS Search} (see Section \ref{sec:ret-hetero}), focusing on datasets and publications in the same system. The assumption followed here is that in an ideal ranking system\footnote{In this paper, by \emph{ranking system} or, \emph{IR system}, we refer to \emph{a system} containing a corpus together with the retrieval model to be used to search on that corpus.}, the retrievability of each indexed item (dataset or other publication) is equally distributed. Likewise, a discrepancy to this assumption may reveal an inequality between the items in a collection caused by the system. By employing a measure of retrievability, we expect to gain further insight into the characteristics of dataset retrieval compared to traditional document retrieval.


\subsection*{Research questions}
Considering the real-life digital library \textit{GESIS Search}, in this paper, we address the following research questions:
\begin{itemize}
    \item \textbf{RQ1:} In the integrated search system with various types of items, can we observe any prior bias of accessibility of documents from a particular type?
    \item \textbf{RQ2:} Can we formalize this type-accessibility bias utilizing the concept of document retrievability?
    \item \textbf{RQ3:} How diverse are the retrievability score distributions in the different categories of documents in our integrated search system?
    \end{itemize}
    
In sum, our contributions are as follows: 1) we utilize the retrievability measure to better understand the diversity of accessing datasets in comparison to publications with real-life queries from a search log; 2) building on retrievability, we propose the measurement of \em usefulness\em, which represents implicit relevance signals observed for datasets and publications. Our understanding of bias follows the argumentation provided in \cite{wilkie2017algorithmic} where bias denotes the inequality between documents in terms of their retrievability within the collection. Bias can be observed when a document is overly or unduly favoured due to some document features (e.g. length, term distribution etc.) \cite{wilkie2014best}.

The rest of the paper is organized as follows.
We first present related work in Section~\ref{sec:rel-work}. 
A formal introduction of the concept of retrievability is presented in Section~\ref{sec:ret}.
The integrated search system \textit{GESIS Search} along with the motivation of our retrievability study is presented in Section~\ref{sec:ret-hetero}.
Section~\ref{sec:exp} discusses the empirical results and analysis of the outcome of the experimentation before introducing the novel concept of usefulness in Section~\ref{sec:usefulness} along with the experimental study of usefulness.
We conclude the paper in Section~\ref{sec:conclusion} highlighting the contributions and findings of the paper. 

\section{Related work}\label{sec:rel-work}
Retrievability as a measure was proposed in \cite{ret-cikm08}. In the work, the authors demonstrate the utility of retrievability on two TREC collections using query based sampling \cite{callan2001query}. 
Retrievability has been primarily used to detect bias in ranking systems. For instance, \cite{samar2018quantifying} employ retrievability to research the effect of bias across time for different document versions (treated as independent documents) in a web archive. Their results show a ranking bias for different versions of the same document.  Furthermore, the study confirms a relationship between retrievability and findability measured by Mean Reciprocal Rank (MRR). They follow the assumption that the lower a document's retrievability score the more difficult it is to find the document.
Another application of the retrievability measure can be found in patent or legal document retrieval which provides a unique use case due to its recall oriented application. In both studies,  \cite{bashir2009analyzing,bashir2009CIKM} look at document retrievability measurements and argue that a single retrievability measure has several limitations in terms of interpretability. In \cite{bashir2009analyzing} they try to improve accessibility measurement by considering sets of relevant and irrelevant queries for each document. In this way, they try to simulate recall oriented users. In addition, they plot different retrievability curves to better spot the gaps between an optimal retrievable system and the tested system. In \cite{bashir2009CIKM} they analyze the bias impact of different retrieval models and query expansion strategies. Their experiments show that clustering based document selection for pseudo-relevance feedback is an effective approach for increasing the findability of individual documents and decreasing the bias of a retrieval system.

 A study on the query list generation phase for determining the measure of retrievability is presented in \cite{bashir2011relationship}. The study addresses two central problems when determining retrievability: 1) query selection and 2) query characteristics identification. It is argued that the query selection phase is usually  performed individually without well-accepted criteria for query generation. Hence their goal is to evaluate how far the selection of query subsets provides an accurate approximation of retrieval bias. The second shortcoming is addressed by determining retrievability bias considering different query characteristics. In their experiments they recognise that query characteristics influence the increase or decrease of retrievability scores. 
 As recognised in \cite{bashir2011relationship}, the majority of retrievability experiments employ simulated queries to determine retrievability. To study the ability of the retrieval measure in detecting a potential retrievability bias using real queries issued by users, \cite{TraubSOHVH16} conducted an experiment on a newspaper corpus. Their study confirms the ability to expose retrievability bias within a more realistic setting using real world queries. A comparison of simulated and real queries with regard to retrievability scores further shows considerable differences which indicates a need for an improved construction of simulated queries. \newline
 
 The concept of usefulness was first introduced in \cite{cole2009usefulness} as a criterion to determine how well a system is able to solve a user's information need. In \cite{hienert2016usefulness}, it has been operationalised within a log-based evaluation approach to determine the usefulness of a  search term suggestion service. In \cite{carevic2018contextualised}, usefulness has been operationalised to determine the effects of contextualised stratagem browsing on the success of a search session. 




Recently, a considerable amount of research has been carried out concerning the characteristics of dataset retrieval. A comprehensive literature review on dataset retrieval practices is provided in \cite{gregory2019searching} focusing on dataset retrieval practices in different disciplines.  Research in this area covers, for instance, the analysis of information-seeking behaviour during dataset retrieval through observations \cite{kramer2021data}, questionnaires and interviews \cite{kern2015there,friedrich2020looking}, and transaction-log studies \cite{kacprzak2017query,carevic2020characteristics}.

In their work in~\cite{kern2015there}, the authors investigated the requirements that users have for a dataset retrieval system. Their findings on dataset retrieval practices suggest that users invest greater effort during relevance assessment of a dataset. They conclude that the selection of a dataset is a much more important decision compared to the selection of a piece of literature. This results in high demands on  metadata quality  during the dataset retrieval. The complexity of assessing the relevance of a dataset is also highlighted in \cite{kramer2021data}.  
Besides topical relevance, access to metadata as well as documentation about the dataset plays a crucial role. A query log analysis from four open data portals is presented in \cite{kacprzak2017query}. Their study indicates differences between queries issued towards a dataset retrieval system and queries in web search. In a subsequent study \cite{kacprzak2018characterising}, the extracted queries are further compared to queries generated from a crowdsourcing task. The intuition and focus of this work is to determine whether queries issued towards a data portal differ from those collected in a less constrained environment (crowdsourcing).


\section{Retrievability}\label{sec:ret}

The \emph{retrievability of a document} in a collection, as defined by Azzopardi and Vinay, is the concept of systematically measuring the possibility of retrieval of the document by \emph{any query}, or in other words, it indicates of how easily the information within the collection can be accessed with a given retrieval model~\cite{ret-cikm08}.
Formally, given a collection $\mathsf{C}$ of documents, and $\mathsf{Q}$, a set of all possible queries for which the answers are expected to be present in $\mathsf{C}$, the retrievability of a document $d$ ($d\in \mathsf{C}$) is defined as:
\begin{equation}\label{eq:ret1}
    r(d) = \sum_{q\in \mathsf{Q}}w_q \cdot f(k(d,q),c)
\end{equation}

In Equation~\ref{eq:ret1}, $w_q$ is the weight of the query $q$, and $k(d,q)$ specifies the rank of document $d$ after querying with $q$ with some retrieval model.
The function $f(k(d,q),c)$ is a boolean function indicative of whether $k(d,q)$ (the rank of document $d$) is within a rank cut-off $c$.
Specifically, the function $f(\cdot)$ returns $1$ if $k(d,q) \leq c$, and $0$ otherwise, as presented in Equation~\ref{eq:fk}.

\begin{equation}\label{eq:fk}
  f(k(d,q),c)=\begin{cases}
    1, & \text{if $k(d,q) \leq c$}.\\
    0, & \text{otherwise}.
  \end{cases}
\end{equation}
Informally, the retrievability score $r(d)$ of a document $d$ can also be realized as the number of queries for which $d$ is ranked within rank-cutoff $c$.

The query weight ($w_q$ in Equation~\ref{eq:ret1}) may depend on the \emph{popularity} of the query which can be treated as a uniform constant for all the queries in $\mathsf{Q}$ to sidestep the effect of query bias (as presented in~\cite{ret-cikm08}); that is, all queries are considered to have equal weights.
The approximated retrievability score of document $d$ will then be a discrete value $x$ indicating the number of queries for which $d$ is retrieved within rank $c$.
Certainly, this is a simplifying assumption and the queries submitted to a search system in practice vary vastly both in terms of \emph{popularity} as well as \emph{difficulty}~\cite{query-dif2006carmel}.
Constructing $\mathsf{Q}$, the set of all queries, expects the utilization of a query log in absence of which a query-based sampling method~\cite{q-sampling} can be applied to randomly populate $\mathsf{Q}$.
Considering a fairly sized collection of documents, there can be infinitely many distinct queries that can be answered by various documents in the collection.
Hence, the construction of $\mathsf{Q}$ based on either query log or random sampling of terms from the collection are only some practical approximations that we can adapt in order to realize the concept of retrievability of documents in a collection.
%


The second factor of the per-query component in Equation~\ref{eq:ret1} is defined to be a boolean function the output of which depends solely on the rank at which document $d$ is retrieved.
Increasing the value of the rank cut-off ($c$) broadens the domain of documents retrieved which will positively influence the retrievability scores of more documents.
Note that, being selected by a retrieval model for some queries does not ensure the relevance of the document which can only be comprehended by human judgements. 

\section{Retrievability in an integrated retrieval system}\label{sec:ret-hetero} 

\begin{figure*}
    \centering
    \includegraphics[scale=0.5]{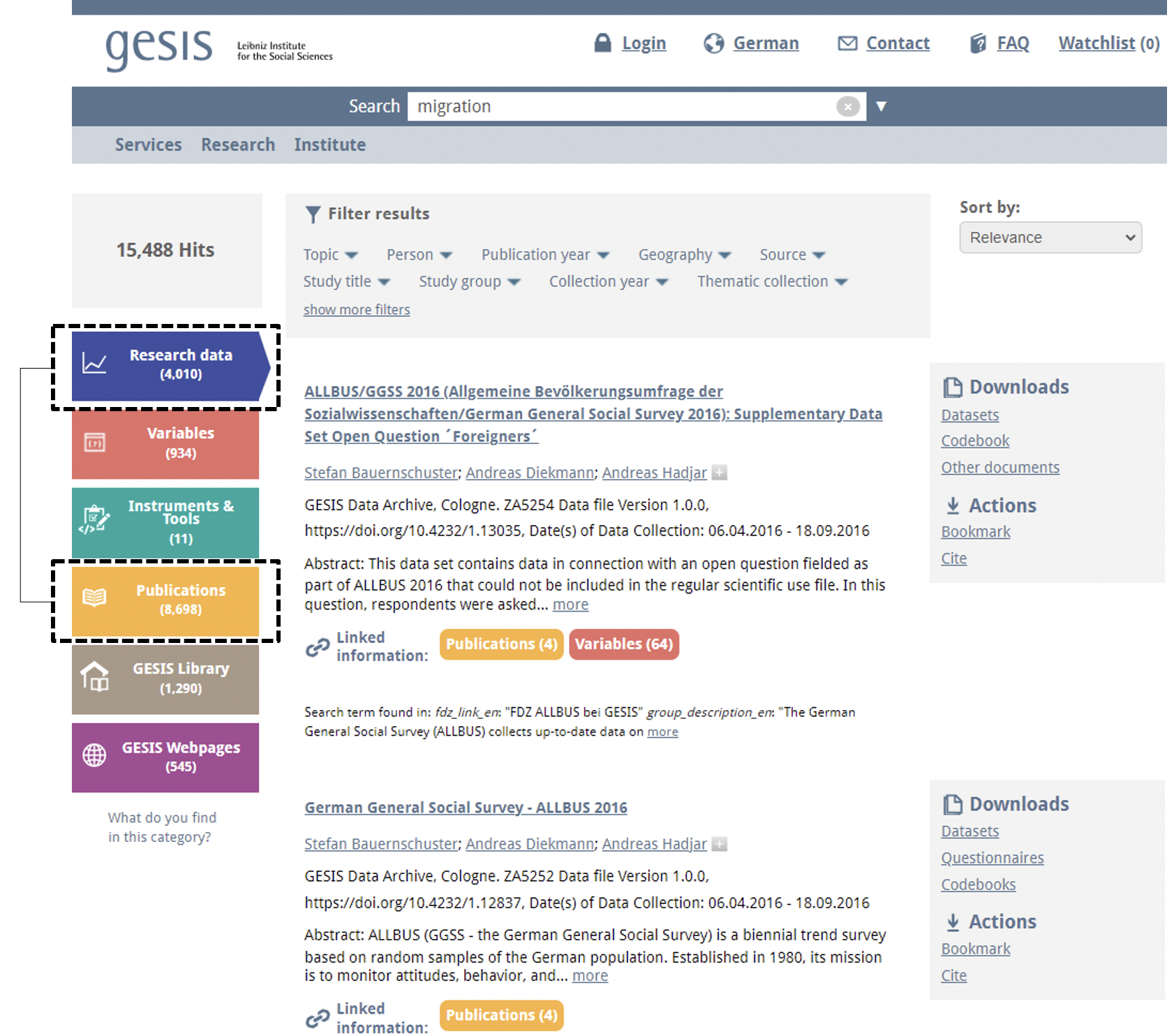}
    \caption{Screenshot of GESIS Search showing result sets for research data and publications.}
    \label{fig:gesis-search-sc}
\end{figure*}

We define an \emph{integrated search system} as a system that searches multiple sources of different types and integrates the output in a unified framework\footnote{This is similar to the concepts of aggregated search~\cite{agreegated-lalmas} or federated search~\cite{federated-arguello}.}.
The retrieval in such a system requires sophisticated decision making considering the various modalities in documents in the collection of data.

Following Equation~\ref{eq:ret1}, the retrievability score of documents is dependent on the other documents in the collection\footnote{Here, we are considering the employed retrieval function as constant.}:
considering a rank-cut $c$, the rank of a document under consideration can be greater than $c$ ($>c$) due to the documents, taking the top $c$ positions, being more relevant or duplicate~\cite{Nikkhoo2011TheIO}.
Another factor that can influence the retrievability score of a document is the popularity; a trendy document will be retrieved multiple times by the users over time.
In case of an integrated search engine, where the documents belong to various categories, some particular types could be having higher chances than the others in terms of being retrieved. 
In general, there can be some disparity in the number of documents of various categories being retrieved which can be a result of popularity bias in the collection. 
This type of popularity bias can impede the satisfaction of the information need of a user, and in turn, can affect the performance of the system.
Traditional methods of query performance prediction~\cite{hauff-qpp,townsend-qpp}, in this scenario, can not be effectively applied.

In this paper, we are going to study the diversity in retrievability scores for different categories of document in the integrated search system \textit{GESIS Search\footnote{https://search.gesis.org} } \cite{hienert2019}.
To the best of our knowledge, this is the first attempt to formalize this type of diversity utilizing the concepts of retrievability.


\section{Experimental Study}\label{sec:exp}

As presented in Section~\ref{sec:ret-hetero}, we use the integrated search system with various categories of documents in this work.
In this section, we start with describing the data that we have used in the work along with different statistics of the data; this will be followed by the experimental evaluation of the study.

\subsection{Datasets}\label{subsec:datasets}

To identify possible differences between publication and dataset retrieval concerning their retrievability, we use the integrated search system \textit{GESIS Search} containing a total of 830K indexed records. A screenshot showing the interface of GESIS search is presented in Figure \ref{fig:gesis-search-sc}.
The indexed records are divided into six categories based on their types, covering more than 118K \emph{publications} and 64K \emph{research datasets} (also referred to as \emph{datasets}). 
Given a query, the system returns six search result pages (SERP) corresponding to each of the categories. 
The segregation of the SERP enables us to study the retrievability of the different types.
In this study, we concentrated on the types \emph{publications} and \emph{datasets} independently. 

In the integrated search system, the interaction of the users with the system is logged and stored in a database. 
A total of more than 40 different interaction types are stored covering, for instance, searches (queries), record views and export interactions etc.~\cite{hienert2019}. 
The export of a record belongs to an umbrella of categories including various interactions such as bookmarking, downloading or citing. These interactions are specifically useful for the application of implicit relevance feedback as they indicate a relevance of a record that goes beyond a simple record view. 
The interaction log of the search system provides the basis for our analysis in Section~\ref{subsec:ana_ret} (and later in Section~\ref{subsec:ana_usefulness}). 
The queries contained in the log (more than 1.2 M) form the basis for determining the retrievability of documents.
This ensures realistic queries in $\mathsf{Q}$ of Equation~\ref{eq:ret1} as opposed to the simulated queries used in~\cite{ret-cikm08} or \cite{TraubSOHVH16}.
The data used in this study was extracted from the interaction log of the integrated search system between July 2017 and December 2021. 
Statistics regarding the extracted interactions utilized in our study can be found in Table~\ref{tab:ext-int}.

\begin{table}[h]
    \centering
\begin{tabular}{ lcp{1.2cm}p{1.6cm}c } 
 \hline
\textbf{Record type} &\textbf{Size}& \textbf{\#queries} (unique) &\textbf{avg. query length} &\textbf{\#exports} \\ 
 \hline
 \multirow{2}{*}{\textbf{Publication}} & \multirow{2}{*}{113,000} & 626,335 (276,553) & \multirow{2}{*}{15.4} & \multirow{2}{*}{8,146} \\
 \multirow{2}{*}{\textbf{Dataset}}& \multirow{2}{*}{64,000} & 602,581 (187,644) &\multirow{2}{*}{20.5} &\multirow{2}{*}{24,310} \\ \hline
\end{tabular}
    \caption{Statistics of the extracted information.}
    \label{tab:ext-int}
\end{table}

\subsection{Measuring retrievability in a collection}\label{subsec:measure-ret}
%
In a traditional relational databases, there can be a one-to-one correspondence between a query and a record. 
In other words, each record stored in R-DBMS can be distinctly selected by a specific query raised to the system.
Also the different queries that can be submitted and addressed by such systems are finitely tractable which depends on the number of records stored.
However, considering the nature of \emph{free-text query}, infinitely many queries can be answered by a traditional information retrieval system through the selection of documents containing the responses.
Also due to the notion of relevance grade (extremely relevant, loosely relevant etc.), and the necessity of having a ranked list, certain documents have more chances of being retrieved than the others.

Considering a traditional document collection $\mathsf{C}$, all the documents are not equally important to a query, hence paving the need to have a ranked retrieval.
Now given a set of all possible queries $\mathsf{Q}$, some documents in $\mathsf{C}$ will be relevant to more number of queries (depending on the topical coverage of the document) than others which can be measured by the concept of retrievability (see Section~\ref{sec:ret}).
Formally with the notion of retrievability, some documents will be having higher $r(d)$ in a collection,
resulting in an unequal distribution of retrievability scores.

Similar types of inequalities are observed in economics and social sciences, and they are traditionally measured using the Gini coefficient or Lorenz curve~\cite{lorenz-gini} which measures the statistical dispersion in a distribution\footnote{Lorenz curve and Gini coefficient are popular in economics to measure of wealth disparity in a community/country.}.

Mathematically, the Gini coefficient (G) of a certain value $v$ in a population $\mathcal{P}$ can be defined as:
\begin{equation}\label{eq:gc}
    G = \frac{\sum_{i=1}^{N}{(2*i-N-1) * v(i)}}{N \sum_{j=1}^{N}{v(j)}}
\end{equation}
where $N$ is the size of the population and $v(i)$ specifies the value of $i^{th}$ item in $\mathcal{P}$.
The value of Gini coefficient in the population will be between $0$ and $1$ and is proportional to the inequality inherent in the population: higher value of $G$ indicates greater disparity and vice versa.
In other words, a value of $G$ equal to $0$ in Equation~\ref{eq:gc} indicates that all the items in the population are equally probable to be selected where as higher values of $G$ specify a bias implying that only certain items will be selected.


\begin{table*}[t]
\begin{tabular}{c rrrr c rrrr}
\hline
\multicolumn{1}{c}{\textbf{Rank}}   & \multicolumn{4}{c}{\textbf{Publication}} & & \multicolumn{4}{c}{\textbf{Dataset}} \\ \cmidrule{2-5} \cmidrule{7-10}
\multicolumn{1}{c}{\textbf{cutoff}} & \multicolumn{1}{c}{\textbf{$\mu$}} & \multicolumn{1}{c}{\textbf{geo-$\mu$}} & \multicolumn{1}{c}{\textbf{$\sigma^2$}} & \multicolumn{1}{c}{\textbf{$\sigma$}} & & \multicolumn{1}{c}{\textbf{$\mu$}} & \multicolumn{1}{c}{\textbf{geo-$\mu$}} & \multicolumn{1}{c}{\textbf{$\sigma^2$}} & \multicolumn{1}{c}{\textbf{$\sigma$}} \\ \hline
10  & 54.86  & 15603.2   & 124.91 & 24.26  &  & 103.27 & 144018.77  & 379.5   & 20.63  \\
20  & 103.95 & 38350.5   & 195.83 & 50.8   &  & 176.63 & 317920.25  & 563.84  & 38.21  \\
30  & 152.49 & 66770.93  & 258.4  & 79.17  &  & 247.01 & 540999.45  & 735.53  & 56.93  \\
40  & 200.36 & 100471.29 & 316.97 & 108.51 &  & 315.71 & 794970.82  & 891.61  & 76.73  \\
50  & 247.66 & 134928.78 & 367.33 & 138.47 &  & 383.59 & 1080000.55 & 1039.23 & 97.57  \\
100 & 477.78 & 352835.83 & 594    & 291.22 &  & 708.76 & 2724762.92 & 1650.69 & 214.74 \\ \hline
\end{tabular}
\caption{The mean (both arithmetic and geometric), variance and standard deviation of the retrievability values when the rank-cutoff is varied.}\label{tab:ret-values-stats}
\end{table*}

\begin{figure*}[t]
  \centering
  \subfloat[Changes in mean of r(d)] {\label{subfig:mean}\includegraphics[width=0.25\textwidth]{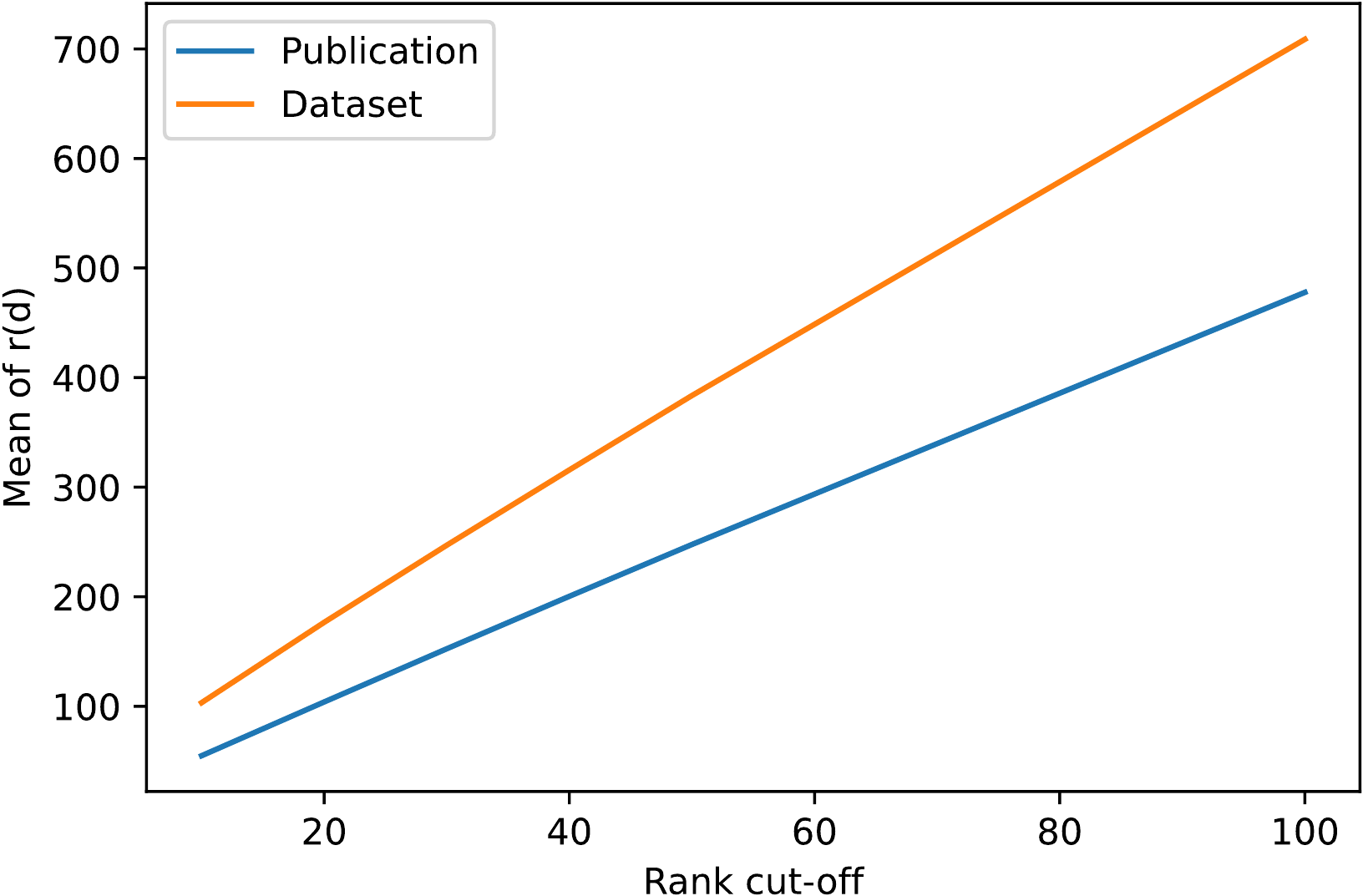}}
  \subfloat[Changes in geometric-mean of r(d)] {\label{subfig:gmean}\includegraphics[width=0.25\textwidth]{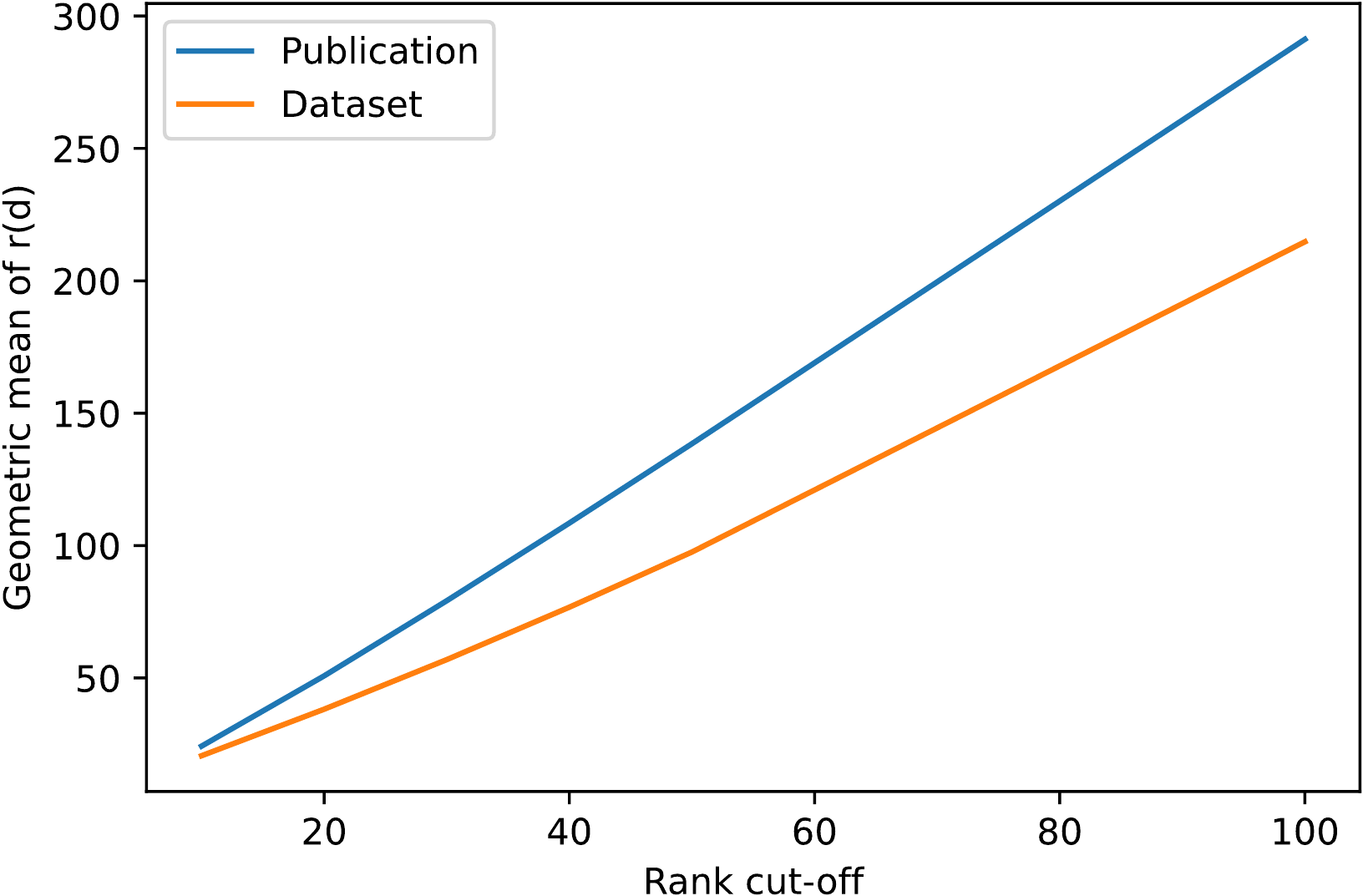}}
  \subfloat[Changes in variance of r(d)] 
{\label{subfig:var}\includegraphics[width=0.25\textwidth]{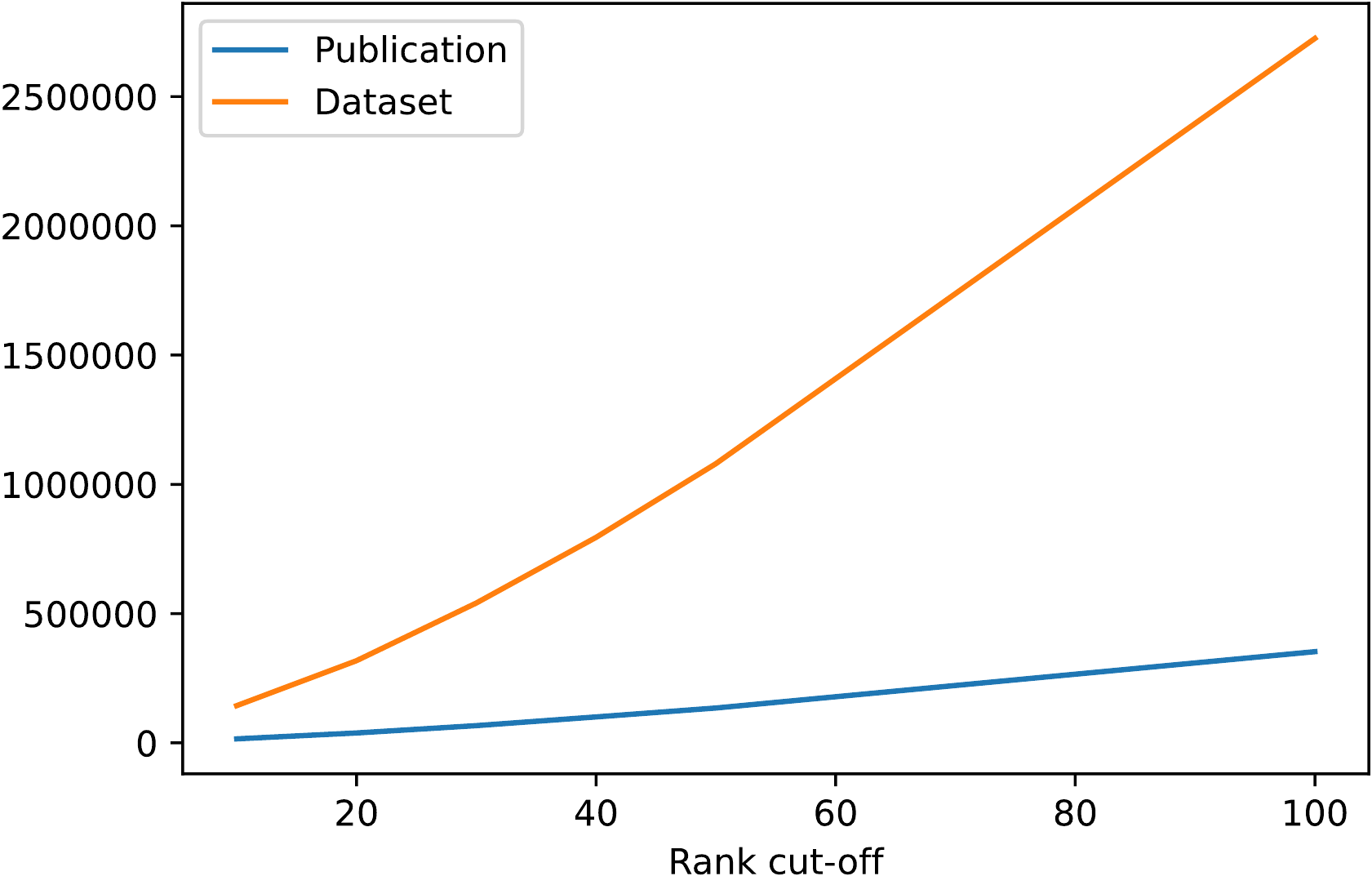}}
  \subfloat[Changes in standard deviation of r(d)] {\label{subfig:std}\includegraphics[width=0.25\textwidth]{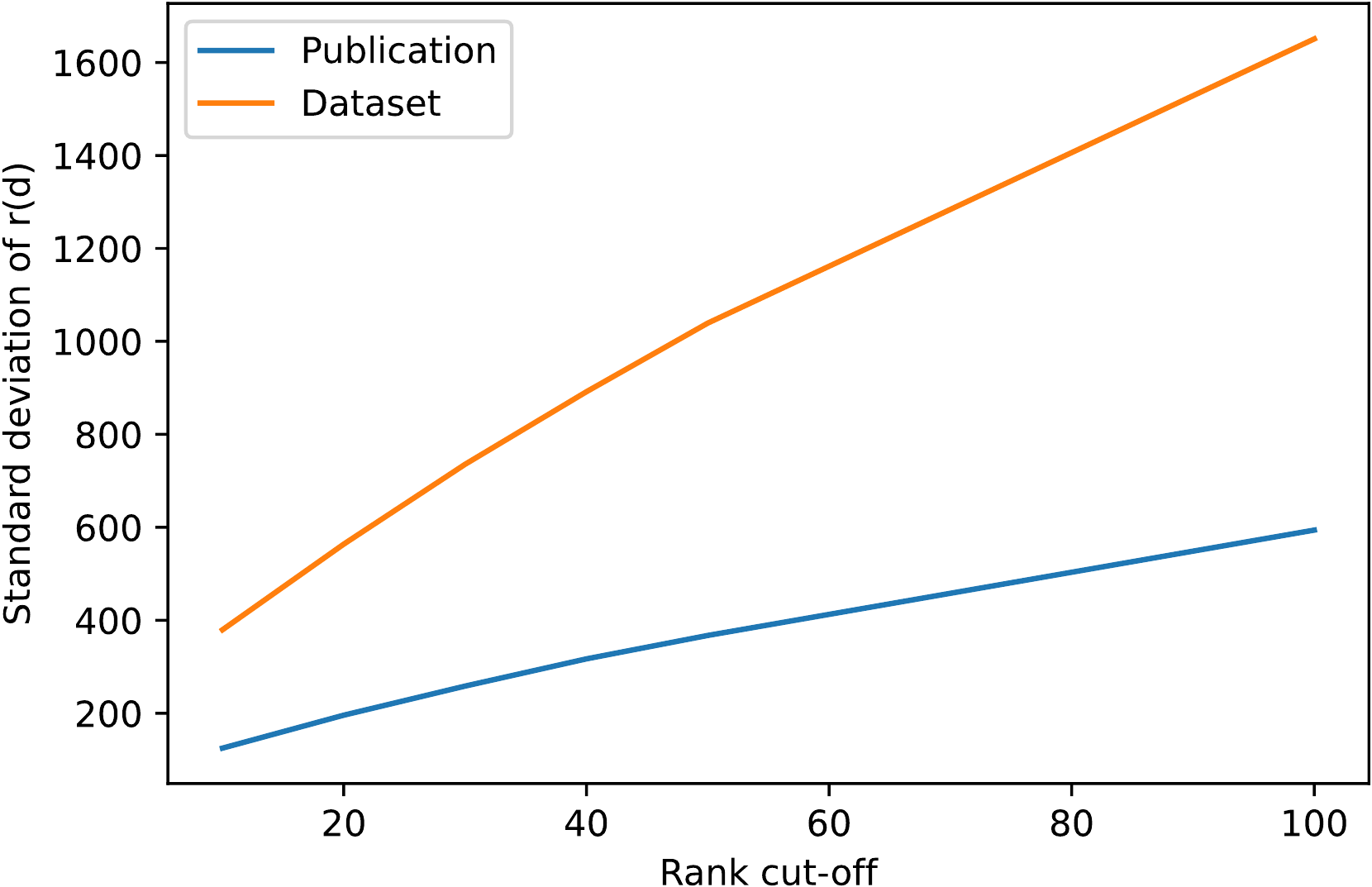}}\\
  \caption{Graphical representation of the change in various statistical measures of the observed distribution of retrievability scores. The mean, geometric-mean, variance and standard deviation of the distribution of retrievability scores of publication (in blue) and dataset (in orange) are presented.
\label{fig:rat-values-stats}}
\end{figure*}

\subsection{Experimentation}\label{subsec:exp}
As explained in Section~\ref{sec:ret}, the retrievability of a document is a measurement of how likely the document will be retrieved by \emph{any} query submitted to the system\footnote{By a system, we are referring to the organization of the collection, along with a retrieval model to be used for retrieval for a given query.}.
Hence, the study of retrievability in a collection of documents requires rigorous retrieval with a set of diversified queries to cover all topics discussed in the collection.
In other words, the retrievability of the documents should be calculated considering all sorts of queries submitted to the system.
However, an infinite number of queries are possible to be answered by a collection for free-text queries.
To cover all the topics, a traditional approximation is to simulate a set of queries randomly, accepting the risk of erratic queries not aligned with the real scenario~\cite{ret-cikm08,TraubSOHVH16}.
With the availability of a query log, the process of query generation can be made more formalized and streamlined to consider the actual queries submitted by real users.
For the study reported in this article, we utilize the query log presented in Section~\ref{subsec:datasets}.

As reported in the earlier study, the retrievability distribution in a collection depends on the employed retrieval model~\cite{ret-cikm08}.
Following the findings by Azzopardi and Vinay, we use BM25 as the retrieval model~\cite{bm25}.
Particularly, we use the implementation available in Elasticsearch\footnote{\url{https://www.elastic.co/}} which use Lucene\footnote{\url{https://www.lucene.apache.org/}} as the background retrieval model.
Following Equation~\ref{eq:ret1}, the retrievability of a document depends on the selection of the rank cutoff value ($c$) - a rank threshold to indicate how deep in the ranked list are we going to explore before finding that document.
Considering the model employed for retrieval and the set of all queries $\mathsf{Q}$ as fixed, $c$ is the only parameter in calculating the retrievability.
For a query $q$, setting a lower value to $c$ will restrict less number of documents being considered retrievable because $f(k(d,q),c)$ will be $1$ only if $k(d,q) \leq c$ (see Equation~\ref{eq:fk}).
Having a higher value of $c$ will allow more documents to be considered retrievable reducing the overall inequality.
In this study, we have varied the value of $c$ in the range $10$ to $100$ in steps of $10$ and have analyzed the observations which is reported in the next section.

\begin{figure*}[t]
  \centering
  \subfloat[Publication] {\label{fig:lc_pub_ret}\includegraphics[scale=0.5]{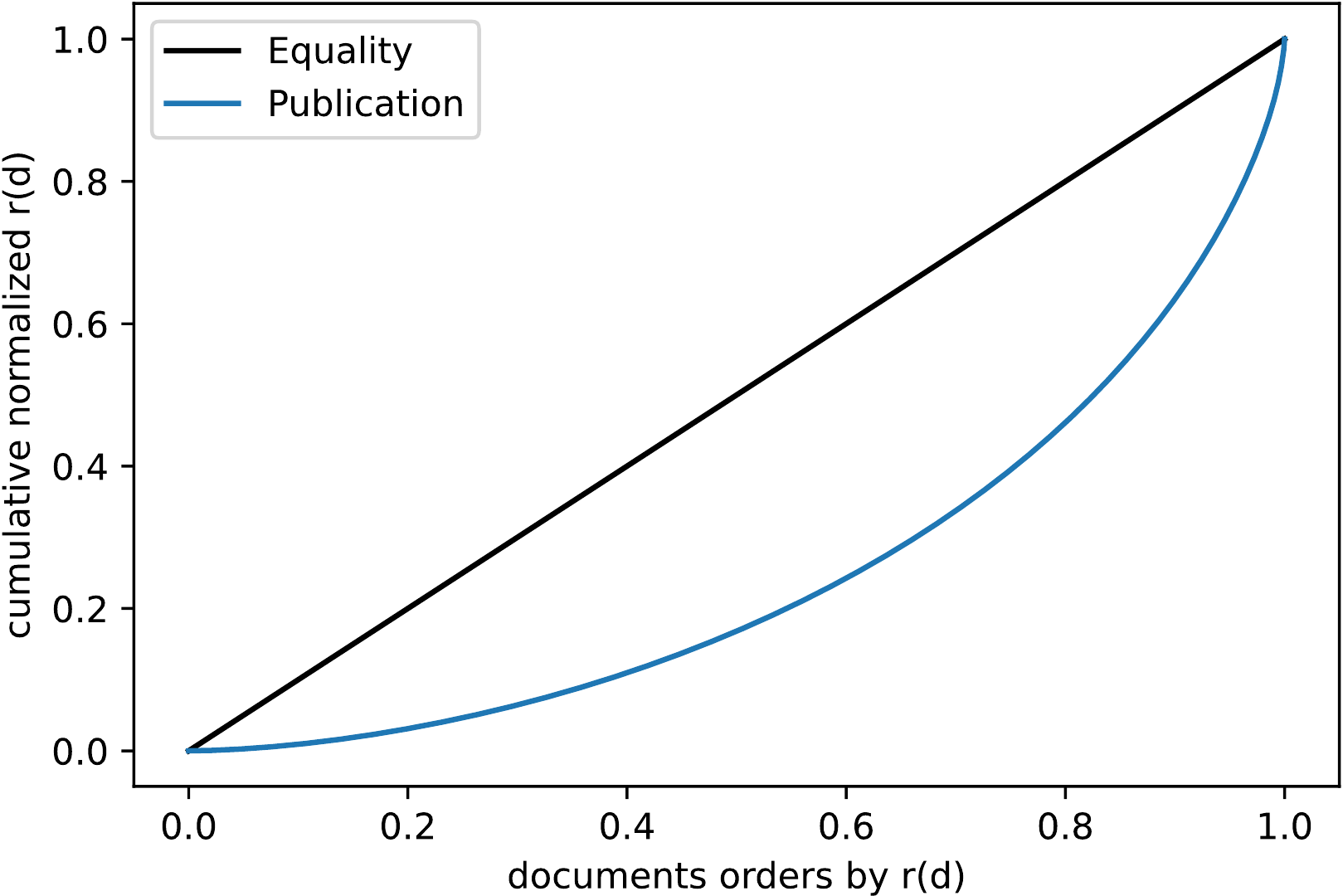}}
  \subfloat[Dataset] {\label{fig:lc_rd_ret}
\includegraphics[scale=0.5]{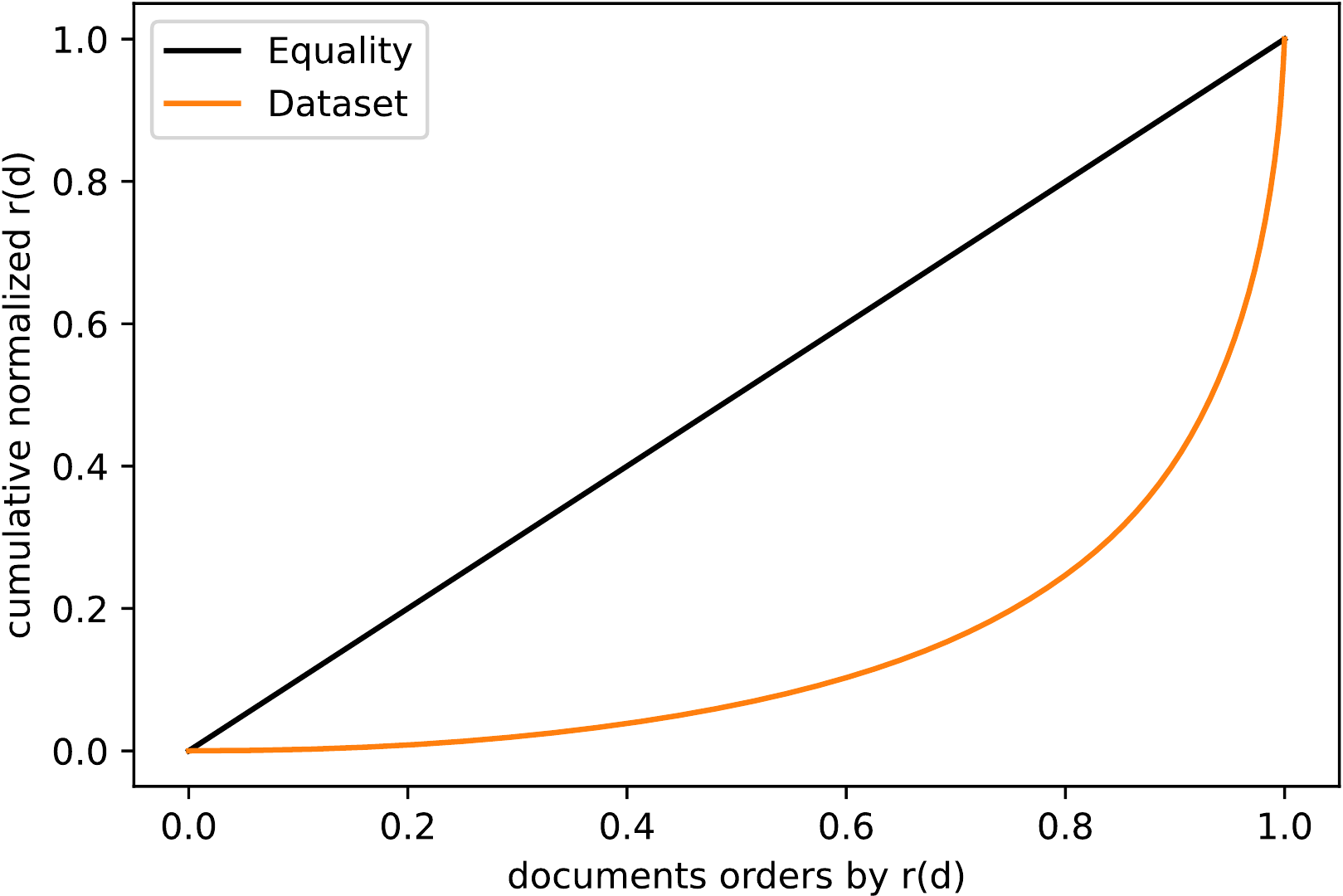}}
    \caption{The Lorenz curve with the retrievability (rank cutoff set to 100). The straight line going through the origin (in black) indicate the equality, that is, when all the documents are equally retrievable.
    }
    \label{fig:ret}
\end{figure*}

\subsection{Observation and analysis} \label{subsec:ana_ret}

We start this section with describing different statistical properties of the retrievability distribution of documents (both publication and dataset) when the value of $c$ is varied.
The mean ($\mu$), geometric mean (geo-$\mu$), variance ($\sigma^2$), and standard deviation ($\sigma$) of the retrievability score distribution on different types (publication and dataset) are given in Table~\ref{tab:ret-values-stats}.
On varying the value of $c$ from $10$ to $100$, we observe a drastic $770\%$ change in the mean retrievability score in case of publication while for datasets, the change in the mean is around $580\%$.
Similar observations are recorded for variance and standard deviation as well when computed using the distribution of $r(d)$ on publication and dataset with different $c$ values.
The variations are presented graphically in Figure~\ref{fig:rat-values-stats}.
From the table, we can conclude that most of the statistical measurements (specifically mean, variance, and standard deviation) are higher for the datasets than publications.
One contrasting observation can be made from Figure~\ref{subfig:gmean} where the variation in geometric mean (\textbf{geo-$\mu$}) is presented.
Unlike the other three statistics, the geometric mean is seen to be higher for publication than dataset across all values of $c$.
Comparing geometric-mean values with the other statistics, we can perceive that for some dataset items, the retrievability values are extensive (for the popular datasets); at the same time, there are number of datasets items with poor $r(d)$ values (for those which are rarely cited/used).
The first category of datasets are contributing to the high average $r(d)$, which is consistent across different $c$ values, while the datasets of the second category cause the geometric-mean to fall.
This is also in line with the observed higher values of variance (and standard deviation) for datasets that indicates diverse $r(d)$ scores.
As opposed to that, we witness a similar behaviour of the publications in terms of both mean and its geometric counterpart.

\begin{figure}
    \centering
    \includegraphics[scale=0.5]{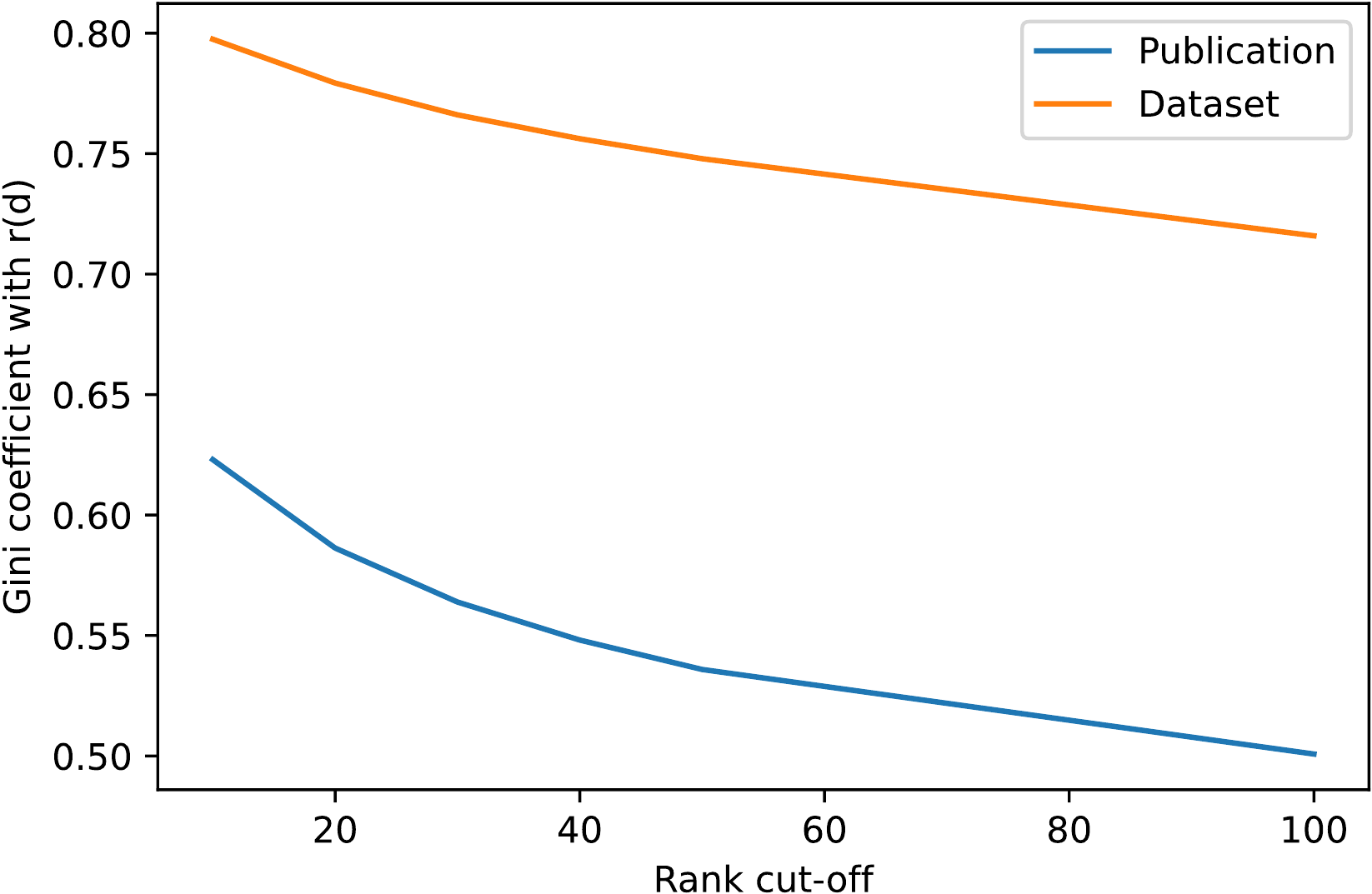}
    \caption{The change in Gini coefficient when the rank cut-off is varied in the range from 10 to 100. The blue line indicates the publication while dataset is specified by the orange curve.}
    \label{fig:change-g_with-c}
\end{figure}

To examine the imbalance in the $r(d)$ scores across the documents of a category, in Figure~\ref{fig:ret} we plot the Lorenz curve with the $r(d)$ scores computed separately for publications and datasets.
The Lorenz curve shows that retrievability of datasets is more imbalanced than publications.
The value of $c$ is set to $100$ in this plot however similar trends are observed with the lower values of $c$ as well.
As discussed in Section~\ref{sec:ret}, the retrievability score of documents escalate with higher values of $c$; consequently, the overall retrievability-balance of the collection also changes positively.
To empirically see the variations in terms of imbalance, Gini coefficients ($G$) attained at different values of $c$ are presented for both the category of documents in Table~\ref{tab:gini-change} which is also graphically displayed in Figure~\ref{fig:change-g_with-c}.
From the table, it can be noticed that the fall of $G$ for publications is more than $19\%$ while there is a $10\%$ change in the $G$ value for datasets.
This confirms that the inequality of documents being searched for is higher in case of datasets than publications.

Additionally, we report the percentage of total documents (respectively of type, publication and dataset) being retrieved while changing $c$ in Table~\ref{tab:gini-change}.
Note that, more than 95\% of documents of type publication are retrieved within the top $10$ retrieved documents; in contrast, just above 76\% datasets are retrievable within the same rank cut-off.
Increasing the value of $c$, it is noticed that more than $98\%$ documents are retrievable within the top $100$ ranked document by all the queries for both publication and dataset.
The significant change in the percentage of retrieved documents of type dataset indicates that searching for datasets is more complex than publications; a deeper ranked list traversal might be essential to find a relevant dataset.

\begin{table*}[t]
\begin{tabular}{c cc c cc cc}
\hline
\multicolumn{1}{l}{\textbf{Rank}}   & \multicolumn{2}{c}{\textbf{Gini coefficient}}                              &  & \multicolumn{4}{c}{\textbf{Retrieved}}       \\ \cmidrule{2-3} \cmidrule{5-6} \cmidrule{7-8}
\multicolumn{1}{c}{\textbf{cutoff}} & \multicolumn{1}{c}{\textbf{Publication}} & \multicolumn{1}{c}{\textbf{Dataset}} & & \multicolumn{1}{c}{\textbf{Publication}} & \multicolumn{1}{c}{\textbf{\%}} & \multicolumn{1}{c}{\textbf{Dataset}} & \multicolumn{1}{c}{\textbf{\%}} \\ \hline
10                         & 0.6230                          & 0.7975               &             & 105795                        & 95.02                  & 48934                         & 76.19                  \\
20                         & 0.5863                          & 0.7793                &            & 109483                        & 98.33                  & 55528                         & 86.45                  \\
30                         & 0.5639                          & 0.7661                 &           & 110413                        & 99.17                  & 58429                         & 90.97                  \\
40                         & 0.5481                          & 0.7562                  &          & 110796                        & 99.51                  & 60025                         & 93.45                  \\
50                         & 0.5359                          & 0.7479                   &         & 110987                        & 99.68                  & 60991                         & 94.96                  \\
100                        & 0.5008                          & 0.7159                    &        & 111269                        & 99.93                  & 62966                         & 98.03                 \\ \hline
\end{tabular}
\caption{Change in Gini coefficient when the rank cut off is increased. Also the number and percentage of documents retrieved of type Publication and Dataset is presented.}\label{tab:gini-change}
\end{table*}

\section{From Retrievability to Usefulness}\label{sec:usefulness}
Usefulness was introduced in \cite{cole2009usefulness} designed initially as a criterion for the evaluation of interactive search systems.
The \emph{usefulness} of a document can be defined as how often the document is retrieved and \emph{exported} (see Section \ref{subsec:datasets})  by the end user. 
Of course the concept of usefulness can only reliably be recognized by relevance judgements submitted by the user for a given query and the relevance of a document may also depend on the perspective of the user which may vary across users.
Without an explicit relevance judgement, the approximation of usefulness or importance of documents can not be reliably accomplished. 
Considering the availability of this information from the query log, we can define the usefulness of a document ($u(d)$) by the following equation:

\begin{equation}\label{eq:usefulness}
    u(d) = \sum_{q\in \mathsf{Q}}w_q \cdot g(d,q)
\end{equation}

In Equation~\ref{eq:usefulness}, the weight of the query ($w_q$) can be defined in a similar way as defined in retrievability (Equation~\ref{eq:ret1}).
The usefulness of a document may also depend on the \textit{difficulty} of the query~\cite{query-dif2006carmel,dif-book2010carmel}\footnote{A query can be considered as \emph{difficult} if the top ranked documents are mostly non-relevant in which scenario, the user has to go deep down the ranked list to get the document addressing the query~\cite{dif-book2010carmel}.}.
A document $d$ should be considered more useful if it is retrieved and consumed following a query $Q$ than any other document, say $d'$ with an associated query $Q'$ which is relatively easier than $Q$ (i.e. $difficulty(Q)>difficulty(Q')$).
Hence, we extend the definition of the weight of the query taking into account a difficulty factor in Equation~\ref{eq:w_q_d}.

\begin{equation}\label{eq:w_q_d}
    w'_q = w_q * h(q)
\end{equation}
where the function $h(q)$ represents the difficulty of the query $q$. 
The function $g(\cdot)$ in Equation~\ref{eq:usefulness} indicates usefulness in terms of relevance of the document $d$ for the query $q$.
Mathematically, $g(\cdot)$ can be defined as follows:
\begin{equation}\label{eq:gk}
  g(d,q) = rel(d,q)
\end{equation}

The function $rel(d,q)$ in Equation~\ref{eq:gk} indicates the relevance of $d$ for the query $q$.
It will work similar to the way function $f(k(d,q),c)$ is defined in Equation~\ref{eq:fk} if a binary relevance, (that is $d$ can be either relevant - $rel(d,q)=1$, or non-relevant - $rel(d,q)=0$ to the query $q$) is considered.

The usefulness of a document cannot be realised until it is retrieved by some retrieval model even though it is relevant.
Given a pair of documents $d_1$ and $d_2$ having the same relevance score, a document, say $d_1$, should be considered more useful than the other ($d_2$) if $d_1$ is ranked higher than $d_2$ in the ranked list produced by a retrieval model. 
Hence, the relevance factor of Equation~\ref{eq:gk} can be extended considering the rank of the document as in the following Equation:
\begin{equation}\label{eq:gk_extended}
  g(d,q) = \frac{1}{k(d,q)} * rel(d,q)
\end{equation}
where the function $k(\cdot)$ indicates the rank of document $d$ after performing a retrieval with query $q$.
Based on Equation~\ref{eq:gk_extended}, we can assert that the usefulness and the rank of the document are inversely proportional to each other; the chance of being consumed by the end user will decrease if the document $d$ is being retrieved at lower rank.
Note that, the function $g(\cdot)$ in Equation~\ref{eq:gk_extended} is dependent on the retrieval model that is used to retrieve $d$ for $q$; the rank-order of a pair of documents in two ranked lists can be different if the corresponding retrievals are performed using different retrieval models. 

\subsection{Experimentation} \label{subsec:exp_usefulness}
As presented and argued earlier in Section~\ref{sec:usefulness}, the signal of document consumption by the user after submitting a query is essential in order to compute the usefulness of the documents.
We utilize the information stored in the interaction-log of the integrated search system as the indication of document consumption by the user.
Particularly, the usefulness is determined on the basis of implicit relevance feedback from the export interactions (see Section~\ref{subsec:datasets}).
A rank cut-off value ($c$) of $10$ is considered; that is, the top $10$ documents of the SERP are considered as viewed/consumed.
Utilizing the difficulty of the query in order to compute the usefulness has been left as part of future work;
the difficulty of the query is kept as constant ($h(q)$ in Equation~\ref{eq:w_q_d} set to $1$) in this study.

\begin{figure*}[t]
  \centering
  \subfloat[Publication] {\label{fig:lc_pub_use}
  \includegraphics[scale=0.5]{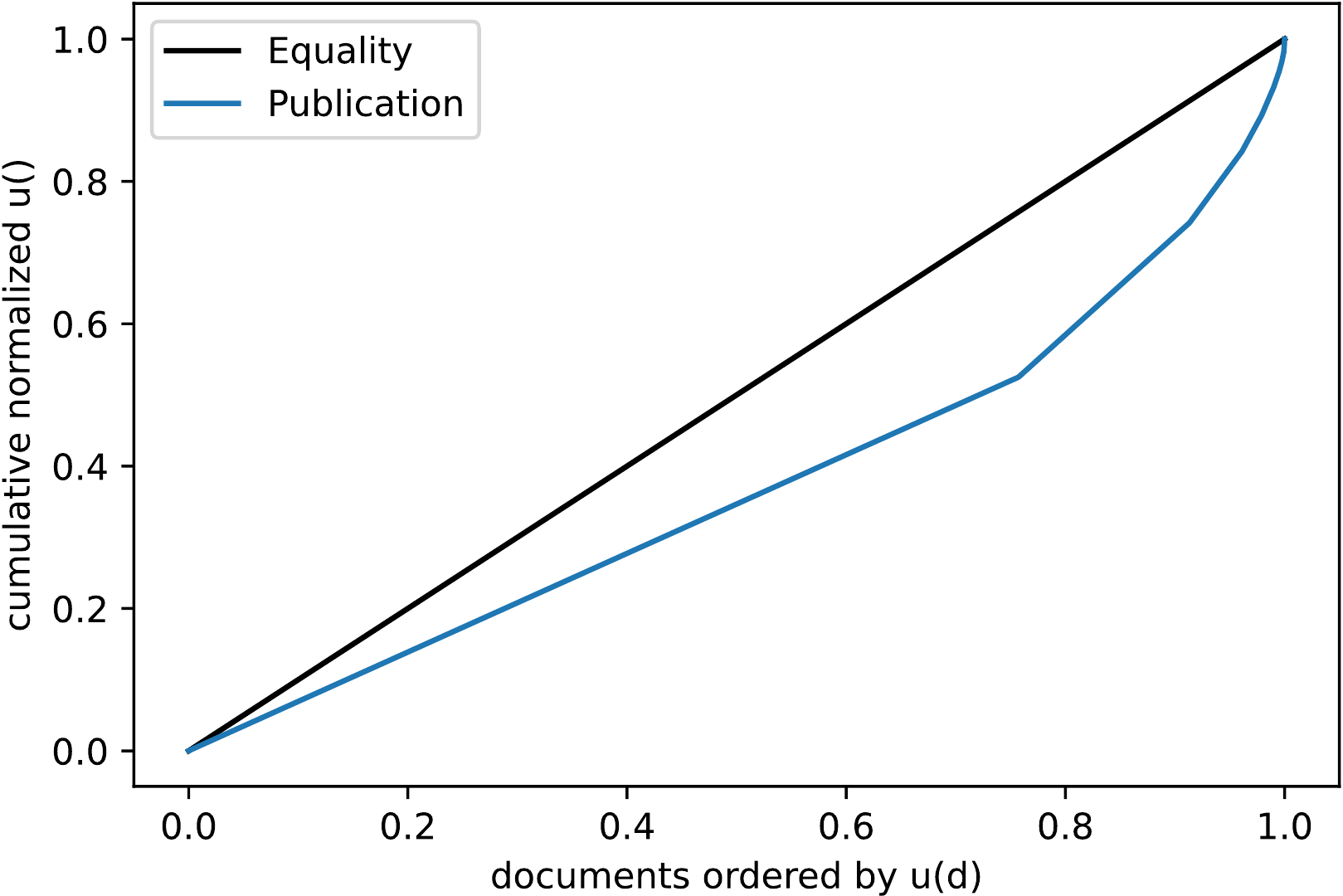}}
  \subfloat[Dataset] {\label{fig:lc_rd_use}
    \includegraphics[scale=0.5]{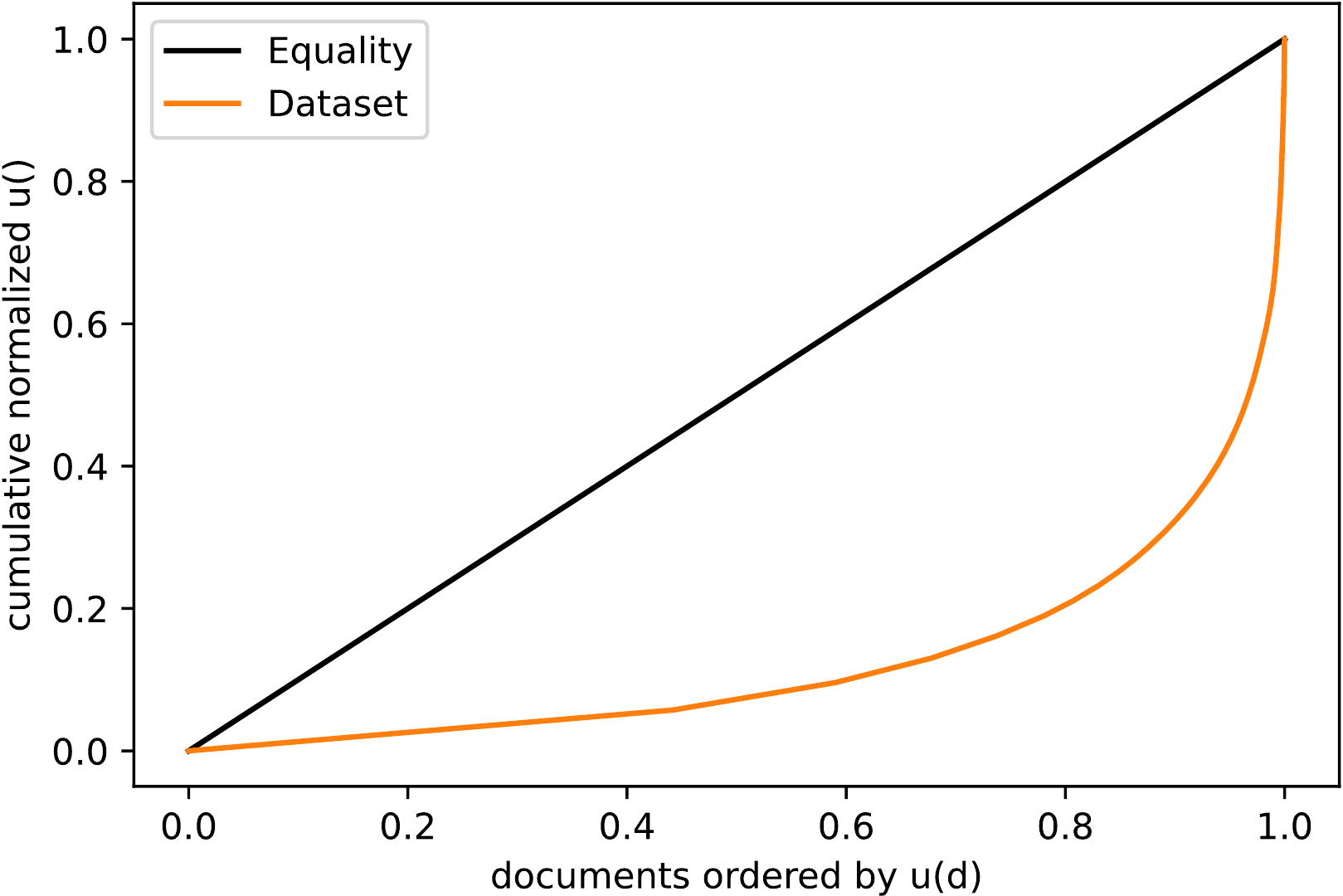}}
    \caption{The Lorenz curve with usefulness. The straight line going through the origin (in black) indicate the \emph{equality}, that is, when all the documents are equally useful.
    The blue line specifies (Figure~\ref{fig:lc_pub_use}) the publication while dataset is indicated by the orange curve (Figure~\ref{fig:lc_rd_use}).
    }
    \label{fig:use}
\end{figure*}

\subsection{Observation and analysis} \label{subsec:ana_usefulness}
The experimental results on the integrated system are graphically presented in Figure~\ref{fig:use} where a pair of Lorenz curves are displayed with the usefulness of the documents of type publication and dataset. 
In Figure~\ref{fig:lc_pub_use}, the usefulness distribution of publications is presented indicating that publications are close to being equally distributed.
In comparison, the similar distribution of datasets (presented in Figure~\ref{fig:lc_rd_use}) is observed to be more skewed with an evident inclination towards certain items being more useful.
The corresponding Gini coefficient of the two distributions (for publications and datasets) is presented in Table~\ref{tab:gini_usefulness}.
The value of $G$ for the usefulness of dataset distribution is seen to be almost three times greater than the publication.
This observation clearly highlights that a few datasets are more useful than the rest, whereas the usefulness distribution of the publications is considerably close to being uniform.

\begin{table}[h]
    \centering
\begin{tabular}{ ccc } 
 \hline
 & \textbf{Publication }& \textbf{Dataset} \\ 
 \hline
 \textbf{Gini coefficient}  & 0.2594 & 0.7466
 \\ \hline
\end{tabular}
    \caption{The Gini coefficient computed with the distribution of usefulness of the publication and dataset. A higher Gini coefficient (upper bound 1.0) indicates an uneven distribution of usefulness.}
    \label{tab:gini_usefulness}
\end{table}

\section{Conclusion}\label{sec:conclusion}

In this paper, we have studied the retrievability of documents from different types belonging to an integrated, real-life digital library, and have observed noticeable differences between documents from the publication and dataset categories.
In response to \textbf{RQ1}, we have seen significant popularity-bias of certain items being retrieved than the others.
Particularly it has been shown that certain items from the dataset category are more likely to be retrieved than the other items of the same category.
In contrast, the retrievability scores of documents from the publication type are more evenly distributed.
For the \textbf{RQ2}, the intra-document selection bias is formalized using the common measures of Lorenz curve and Gini coefficient. 
In response to \textbf{RQ3}, we have observed that the distribution of document retrievability is more diverse for the dataset as compared to publications.
This can be again attributed to the popularity bias of certain items in the dataset category.

Additionally, we have proposed a measurement of \emph{usefulness} of documents based on the signal of document consumption by the users after submitting a query to the system.
From the experimentation, we have noticed that the usefulness of documents from the publication category, in general, is substantially greater than the usefulness of documents from the datasets.
Comparing the intra-category usefulness, a vast diversity is reported for datasets which signifies that certain documents are considerably more \emph{trendy} in terms of being retrieved and satisfying the information need in case of searching in datasets than publications.


\paragraph{\textbf{Acknowledgement}}\label{sec:acknowledgement}
This work was funded by DFG under grant MA 3964/10-1, the Establishing Contextual Dataset Retrieval - transferring concepts from document to dataset retrieval (ConDATA) project, \url{http://bit.ly/Condata}.


\bibliographystyle{ACM-Reference-Format}
\bibliography{refs}


\end{document}